\documentclass[twocolumn,showpacs,preprintnumbers,pra,amsmath,amssymb]{revtex4}

\usepackage{graphicx}
\usepackage{bm}
\begin{document}
\title{Optical Activity and Mirror-Symmetry}

\author{Won-Young Hwang
\footnote{Email: wyhwang@jnu.ac.kr}
}
\affiliation{Department of Physics Education, Chonnam National University, Gwangju 500-757, Republic of Korea}
\begin{abstract}
A misconception that non-chiral molecules have no optical activity at all is widespread. However, at molecular level even non-chiral molecules have optical activity. Optical activity of a non-chiral molecule is canceled out by that of another molecule in its mirror image in normal liquids. We describe the canceling mechanism by using mirror-symmetry of physical laws without resorting to detailed formulas. The description will be helpful for overcoming the misconception. Optical activity can be understood from the opposite viewpoint by the description. Aligned non-chiral molecules have optical activity.
\end{abstract}
\maketitle
\section{Introduction}
Chiral molecules are those which cannot be transformed into its mirror image only by spatial rotation. Chiral molecules are optically active, namely they rotate polarization plane of light \cite{Hec02,Jen50,Con37,Mas68}. Non-chiral (or achiral) molecules in liquid where molecules are randomly oriented show no optical activity. However, widespread is a misconception that non-chiral molecules have no optical activity at all. However, even non-chiral molecules have optical activity at molecular level \cite{Gib82,Cla06,Isb07,Bar04}. The reason why liquids composed of non-chiral molecules show no optical activity is that effect of a molecule is canceled out by that of another molecule in its mirror image.

The purpose of this paper is to overcome the misconception by introducing a description of the canceling mechanism. The description neatly makes use of only mirror-symmetry of physical laws without resorting to detailed formulas. Thus it has also pedagogic value.


\section{Rotatory power cancels by mirror-symmetry}
Suppose that light polarized in $x$ direction propagates in $+z$ direction passing through a molecule, which may be chiral or non-chiral, in a fixed direction (Fig.1).  Electro-magnetic field of incident light induces current in the molecule which then generates secondary electro-magnetic wave, or scattered light. (Here we consider classical model as described in Ref. \cite{Hec02}. Our argument is also valid in quantum case that is also mirror-symmetric.)
\begin{figure}
\includegraphics[width=9cm,height=7cm]{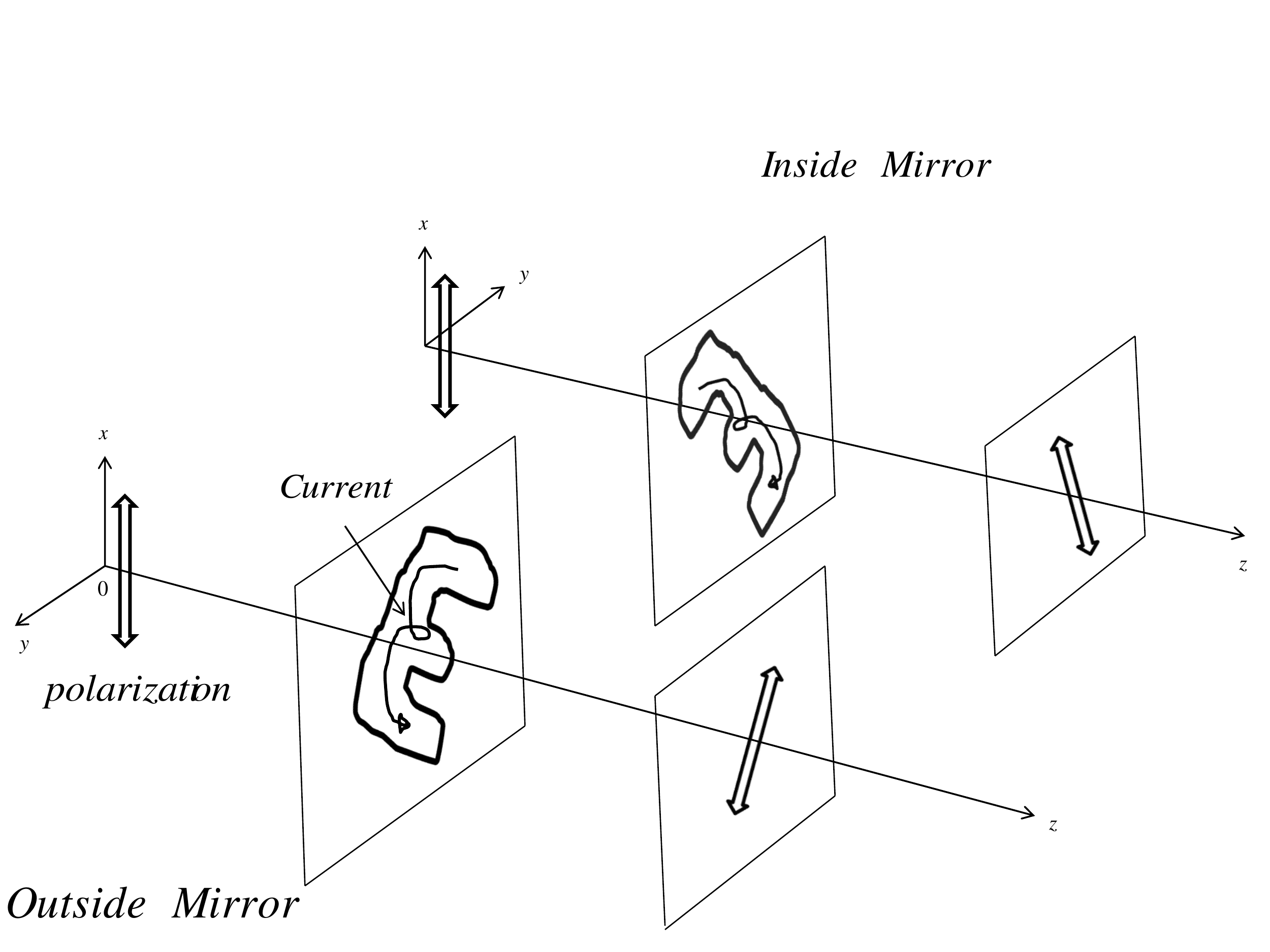}
\caption{Light polarized in $x$ direction propagates in $+z$ direction passing through a molecule in a fixed direction.  Electro-magnetic field of incident light induces current in the molecule which then generates secondary electro-magnetic wave. As a result, polarization plane rotates. What happen in the mirror can also happen in real world by space-inversion symmetry. Rotatory power of a molecule is negative of it's mirror image's.}
\label{Fig-1}
\end{figure}
Combination of the scattered light and remaining original one is resultant light. Thus polarization of scattered light influences that of resultant one. If polarization of the scattered one is rotated, then polarization of resultant one should also change in general.


Let us see how the effect on polarization direction, the rotatory power, of single molecules is canceled out. Let us consider a mirror image of the event we have considered. The same physical laws applies to event in the mirror. Therefore, the mirror image is what really happens if we had performed the experiment with proper initial conditions. If light polarized in $x$ direction propagates in $+z$ direction passing through a molecule fixed in a direction as that of the mirror image, the plane of polarization of resultant light would be that of mirror image. Rotation angle of polarization direction is negative of the original one. Here we don't have to care detailed mechanism. The mirror-symmetry simply implies that rotatory power of a molecule is negative of it's mirror image's.

By definition a non-chiral molecule in a fixed direction can be transformed to its mirror image by rotation. Molecules in normal liquid are randomly oriented. Therefore, on average, effect on polarization is canceled out. Let us discuss it in detail. To describe orientations of a molecule is the same as to describe that of a (three-dimensional perpendicular) coordinate system. Thus, as in the case of Euler angles, three variables are required. Let us denote an orientation of a molecule by a set of three variables, $\vec{X}$. Here $\vec{X}$ can be Euler angles but may be those from other angle-system to describe the orientation. Suppose that light polarized in $x$ direction is rotated by an amount $f(\vec{X})$ after passing through a molecule in an orientation $\vec{X}$. What is implied by the mirror symmetry is that for any $\vec{X}$ there exists an $\vec{X}^{\prime}$, which is the mirror image, such that
\begin{equation}
f(\vec{X}^{\prime})= -f(\vec{X}).
\label{1}
\end{equation}
A non-chiral molecule in an orientation $\vec{X}$  can be transformed to its mirror image in an orientation $\vec{X}^{\prime}$ by simple rotation. Because molecules in normal liquids are randomly oriented, on average, effect of the molecule in an orientation is canceled out by that of the molecule in a mirror-image orientation. Namely
\begin{equation}
\sum_{\mbox{all } \vec{X}} f(\vec{X})= 0.
\label{2}
\end{equation}
(Because orientations are continuous variables we should discuss it by using integrals with proper measure. However, Eq. (2) is quite clear intuitively. Thus we omit the discussion to avoid complications.)
What we have calculated is ensemble average of the molecule's effect. Exactly speaking, what we need is an average along a light path. However, we can see that the ensemble average is the same as the path-average, because random distribution is invariant with respect to coordinate-system rotation.
\section{discussion and conclusion}
In the case of chiral molecules, a mirror image cannot be obtained by simple rotation of the original one by definition. Hence Eq. (\ref{1}) is not obtained and the ensemble average in Eq. (\ref{2}) is not zero in general. This is an explanation from the opposite viewpoint about why liquids of chiral molecules has optical activity.

We can expect that if molecules are aligned in a fixed direction, even non-chiral molecules have rotatory power \cite{Bar04,Isb07}. In this case,
polarization direction changes after passing through each molecule while molecules are in the fixed direction. Thus
relative angle between polarization direction of incident light and orientation of molecule changes and thus amount of rotation due to each molecule is not the same. Suppose that for a certain relative angle between polarization direction of incident light and orientation of molecule, the amount of rotation due to a molecule is zero. Once the polarization direction reached at the stationary angle, the polarization direction will not change. In this aligned molecules case, we can see that the effect on polarization direction will not be proportional to length of light path, differently from normal case. This is what happens in aligned liquid crystal.

It is proposed in Ref. \cite{Isb07} to measure optical activity of water molecules aligned in ice XI, a certain phase of ice. However, water molecules can also be (partially) oriented by strong electric field. It will be worthwhile to observe optical activity due to the partial alignment.

A misconception that non-chiral molecules have no optical activity at all is widespread. However, at molecular level even non-chiral molecules have optical activity. The reason why normal liquids composed of non-chiral molecules show no optical activity is that effect of a molecule is canceled out by that of another molecule in its mirror image. We described the canceling mechanism by using mirror-symmetry of physical laws without resorting to detailed formulas. This is an explanation for optical activity from the opposite viewpoint. Aligned non-chiral molecules have rotatory power, which is not proportional to length of light path.

I am grateful to Trond Saue who informed me about existence of Refs. \cite{Gib82,Cla06,Isb07,Bar04}. This study was supported by Basic Science Research Program through the National Research Foundation of Korea (NRF) funded by the Ministry of Education, Science and Technology (2010-0007208).

\end{document}